\documentclass[journal,10pt]{IEEEtran}
\usepackage{ifpdf}
\usepackage{amsmath}
\usepackage{amsopn}
\usepackage{subcaption}
\usepackage{algorithmic}
\usepackage{array}
\usepackage{url}
\usepackage{booktabs}
\usepackage{multirow}
\usepackage{float}
\usepackage[table,xcdraw]{xcolor}
\usepackage{gensymb}
\usepackage{textcomp}
\usepackage{graphicx}
\graphicspath{ {./images/} }
\usepackage{amssymb}
\usepackage{caption}
\usepackage{calc}
\usepackage{fancyhdr}
\usepackage[
backend=biber,
style=ieee,
citestyle=ieee,
sorting=none
]{biblatex}
\begin{document}

\title{Estimating State of Charge for xEV batteries using 1D Convolutional Neural Networks and Transfer Learning}

\author{Arnab~Bhattacharjee,~Ashu~Verma,~\IEEEmembership{Senior Member,~IEEE},~Sukumar~Mishra,~\IEEEmembership{Senior Member,~IEEE},~Tapan~K.~Saha,~\IEEEmembership{Fellow,~IEEE}% <-this % stops a space
\thanks{Arnab Bhattacharjee is enrolled as a Ph.D. student in Electrical Engineering at the University of Queensland-IIT Delhi Academy of Research. email:  Arnab.Bhattacharjee@uqidar.iitd.ac.in}%
\thanks{Ashu Verma is associated with the Centre for Energy Studies, Indian Institute of Technology, Delhi, India, 110016. email: averma@ces.iitd.ac.in}%
\thanks{Sukumar Mishra is associated with the Electrical Engineering Department, Indian Institute of Technology, Delhi, India, 110016. email: sukumar@ee.iitd.ac.in}%
\thanks{Tapan K Saha is at the School of Information Technology and Electrical Engineering in the University of Queensland, Brisbane, Australia. email: saha@itee.uq.edu.au}}%

%\markboth{IEEE Transactions on Vehicular Technology,~Vol.~XX, No.~XX, XXX~2020}{}


\maketitle
\begin{abstract}
In this paper we propose a one-dimensional convolutional neural network (CNN)-based state of charge estimation algorithm for electric vehicles. The CNN is trained using two publicly available battery datasets. The influence of different types of noises on the estimation capabilities of the CNN model has been studied. Moreover, a transfer learning mechanism is proposed in order to make the developed algorithm generalize better and estimate with an acceptable accuracy when a battery with different chemical characteristics than the one used for training the model, is used. It has been observed that using transfer learning, the model can learn sufficiently well with significantly less amount of battery data. The proposed method fares well in terms of estimation accuracy, learning speed and generalization capability.
\end{abstract}


\IEEEpeerreviewmaketitle

\section{Introduction}\label{sec:intro}

\IEEEPARstart{B}{attery} powered electric vehicles (EVs) are rising in demand as the future of the automobile sector. They are being seen as the most immediate green alternatives to conventional fossil fuel powered vehicles. However,  the  use  of batteries  as  the  source  of  power  for  electric  vehicles  poses  a very unique problem, one that their fuel powered counterparts never  came  across – the lack of a fuel gauge like system to provide information about the amount of energy left to drive the vehicle. 

The state of charge or SoC is a quantity which provides the information about the remaining amount of charge in a battery as a ratio to its nominal capacity, and hence indirectly provides information about the remaining amount of energy in the battery. Unfortunately, it cannot be directly measured and thus has to be estimated indirectly from the observable battery parameters and variables. SoC estimation is thus of paramount importance (\cite{5609223}) owing to its applications in range estimation of an EV, initial charge estimation for charging operations, cell energy balancing operation, demand response and so on. Over the years, a wide range of battery SoC estimators have been proposed in the literature and those can be broadly categorized into four types: lookup table method, ampere hour integral method, model based estimation method and model free estimation method.

The most basic form of SoC estimators are based on a lookup table approach. A lookup table is built by identifying a relation between the SoC and some other measurable battery parameters like Open circuit voltage (OCV) or the AC impedance. During battery operation, the value of the measurable parameter is obtained and the SoC is looked up from a table (\cite{RODRIGUES200012},\cite{86c5fe76266347ef896b1ae0992c0d87}) . Such methods are not scalable and only provide highly approximate values of SoC. Moreover, the observable battery parameters that are mostly used for the look up table correspond to battery static characteristics like OCV. Thus in order to measure these parameters, the battery needs to be put at rest for a long time to reach steady state.

If the initial SoC is known and the current can be measured precisely, the amount of charge lost or gained by the battery can be calculated by simply integrating the current over the time of battery operation. This simple approach known as the ampere hour integral method requires that the initial SoC is known and that the error or disturbances in current measurement is negligible. Also the total charge capacity of the battery needs to be re-calibrated with time owing to battery aging effects. This method is suitable for laboratory applications \cite{5609223}.

Model based estimation techniques use a mathematical model of the battery to form their state equations. Following this, adaptive filters and non linear estimation algorithms can be used to determine or estimate the internal states of the battery. Various non linear state estimation algorithms and filters have been used in literature for SoC estimation. Kalman filters and its various types have been widely used across literature for SoC estimation((\cite{PLETT2004277}),\cite{5739545},\cite{6323045},\cite{7984834}) owing to their robustness, accuracy of estimation and ease of implementation. The estimation accuracy of these model based estimators heavily depend on the goodness of the battery models on which they are built. These models, however, are almost always based on some underlying simplifying assumptions necessary to make them tractable. 
Also, while a huge amount of literature exists for methods that deal with noise in data during state estimation, making the battery state estimation robust, the consideration of noise while estimating the parameters of the battery model from collected data is often missing. As has been shown in \cite{WEI2020114932}
, the use of the vanilla least squares regression method or any of its variants to identify the parameters of a battery model under noisy conditions, without compensating for the ensuing noise induced biases, leads to an underestimation of the actual parameters. Using such a model without compensating for the biases induced by noise into its parameters will lead to faulty state estimation. \cite{WEI2020114932}, \cite{8951446} talk about noise immunity in model parameterization for battery models and propose methods that eliminate such noise induced bias in model parameters.

In situations where the entire mathematical model of the battery is unknown or fails to account for the uncertainties in the system, model free approaches are the most preferred. As the name suggests these methods use learnable non linear functions to adaptively identify the battery model from the observable battery variables, like terminal voltage, input current, cell temperature, etc. and hence are capable of estimating the internal states of the battery directly from data. Machine learning and deep learning algorithms fall under this category. They are robust to noise, provide low values of estimation error and are highly scalable and easily deployable.

Different types of neural networks have been used previously in literature for the purpose of SoC estimation. \cite{4138019} used fuzzy neural networks and learning controllers for estimation purposes. A feed forward neural network was used for SoC estimation in \cite{CHEMALI2018242}, \cite{8269299} and a backtracking search algorithm was employed for hyperparameter tuning in \cite{8269299}. In \cite{7110398}, a robust sliding mode observer was used for the estimation purpose and an RBF neural network was used for adaptively developing an upper bound on the uncertainties. 

Considering the temporal nature of battery observable data, numerous deep learning papers have come up where different forms of Recurrent Neural Networks have been proposed for SoC Estimation. \cite{7948779} used dynamically driven RNNs for online estimation of SoC. \cite{8240689} used Long Short Term Memory (LSTM) unit networks for SoC estimation. \cite{8703041} used a hybrid Variable Auto-Regressive Moving Average(VARMA)- LSTM combination in order to separately estimate the linear and non linear components of current in order to estimate SoC. \cite{Li_2019}, \cite{8698286} used Gated Recurrent Unit (GRU) networks for estimation purposes. 

Convolutional Neural Networks (CNNs) are another powerful class of deep learning algorithms used for image and time series data processing. In \cite{8754752}, the authors use a hybrid CNN-LSTM network to estimate SoC where the spatial features from current input are extracted using the CNN while the temporal features from past inputs are extracted using the LSTM network. To the best of our knowledge, the use of CNNs in a standalone fashion for SoC estimation has not been previously observed in literature. 

Deep learning algorithms, although preferred for their accuracy in estimation, suffer from the limitation of being task specific. Transfer Learning refers to a class of algorithms that aim at solving this problem. \cite{8790543} used transfer learning for improving the generalization capability of their LSTM based SoC estimator. However, they used very shallow networks leading to high estimation errors.

This paper proposes a model free SoC estimation technique based on a deep wide one dimensional(1D) CNN . The width in the CNN allows for the identification of long and short term trends in the battery data. Moreover, to ensure higher generalizability and data efficiency, a transfer learning framework is proposed.

The key claims of this paper are highlighted below:\\
\begin{enumerate}
    \item The 1D CNN based SoC estimator performs better than or at least at par with the existing state of art SoC estimators in terms of estimation error.
    \item The time of offline training required before deploying the proposed model is lesser than the existing state of art machine learning based SoC estimators. The time further decreases when transfer learning is used.
    \item Using transfer learning, the model learns to predict on a target battery dataset accurately after being trained with a substantially less amount of data than is required when training without transfer learning.
\end{enumerate}

The rest of the paper is organised in the following manner. Section \ref{sec:Proposed method} consists of a brief description of the SoC estimation problem formulation and introductions to the convolutional neural network and transfer learning algorithms used in the paper. The datasets are elaborated in section  \ref{sec:data desc}. Section \ref{sec:exp setup} consists a description of the experimental setup and section \ref{sec:results} contains the experimental results and inferences. The paper is concluded in section \ref{sec:conclusion}.

\section{Proposed Method}\label{sec:Proposed method}
\subsection{Problem Statement}\label{subsec:Prob statement}
The problem of estimating the state of charge of a battery is modelled as a time series analysis problem. The SoC at timestep $k$ is formulated as a function of the observed battery parameters, i.e., the current, voltage and the battery temperature, measured till the timestep $k$. Here, we first model the SoC in the following way:

\begin{equation}
    \begin{aligned}{}
        \widehat{SoC}_k &=\mathnormal{f(\phi_k, \phi_{k-1},...,\phi_1)}
    \end{aligned}
    \label{eq:SoC_1}
\end{equation}{}
where, $\mathnormal{\phi_k =[V_k,I_k,T_k]^T}$ and $\mathnormal{k = 1,...,t_{end}}$

where $V_k$, $I_k$ and $T_k$ are the voltage, current and temperature values measured at timestep $k$, $t_{end}$ is the final timestep and $f$ is a non linear mapping from the feature space to the label space.

However, with the functional model defined in equation \ref{eq:SoC_1}, depending on the value of $k$, we would have to vary the input size or dimensions of the data driven estimator which can lead to  difficulties during training. Secondly, with an increasing value of $k$ the capacity of the CNN model will have to be increased to ensure it captures relevant features from data, thus making it computationally burdensome. Keeping these in mind, we propose to model the SoC at the $k$-th timestep as a function of the observed battery parameters ranging from $k$ to $k-t_w+1$ timestep, i.e., we take into account a fixed range of historical data as input to estimate the current SoC value. The effects of varying the time horizon/ range/ window, $t_w$, has been studied in the later sections. Now the modified SoC model will look like this:

\begin{equation}
    \begin{aligned}{}
        \widehat{SoC}_k &= \mathnormal{f_{ \theta}( \phi_k,  \phi_{k-1},..., \phi_{k-t_w+1})} \forall \mathnormal{k \geq t_w \geq 0 }
    \end{aligned}
    \label{eq:SoC_2}
\end{equation}{}
with the symbols holding their usual meaning.

With this functional model, we now formulate an optimization problem whose solution would give us the optimal mapping from the feature space, consisting of the time varying observed battery parameters to the label space, consisting of the SoC. Note that in equation \ref{eq:SoC_2} we have parameterized $f$ by $\theta$ and in doing so, we convert the computationally expensive problem of optimization in a functional space to a significantly simpler problem of parameter optimization. In our proposed method, a 1D convolutional neural network is used as $f$  where  $\theta$ represents its weights and biases such that  $\theta \in \Theta_C$ where  $\Theta_C$ refers to the CNN parameter space. We can now formulate an optimization problem as shown below:
\begin{equation}
    \begin{aligned}
        \theta ^* &= \mathop{\textrm{argmin}}_{\theta \in \Theta_C} G(|SoC_k - \widehat{SoC}_k|, \forall k= t_w,...,t_{end}) 
    \end{aligned}
    \label{eq:theta_argmin}
\end{equation}
where $G$ is a distance function, taking as arguments, the absolute difference between the estimated and actual SoC values at each timestep. $G$ can be a Mean Squared Error (MSE) function as shown in equation \ref{eq:MSE}, a Mean Absolute Error function(MAE) as shown in equation \ref{eq:MAE} or a Maximum Absolute Error (MAX) function as shown in equation \ref{eq:MAX} . However $G$ is not restricted to these functions only.

\begin{equation}
    \begin{aligned}
        MSE &=  \frac{1}{L} \sum_{k=t_w}^{t_{end}}(|SoC_k-\widehat{SoC}_k|^2)
    \end{aligned}
    \label{eq:MSE}
\end{equation}{}

\begin{equation}
    \begin{aligned}
        MAE &=  \frac{1}{L} \sum_{k=t_w}^{t_{end}}(|SoC_k-\widehat{SoC}_k|)
    \end{aligned}
    \label{eq:MAE}
\end{equation}{}

\begin{equation}
    \begin{aligned}
        MAX &= max\{(|SoC_k-\widehat{SoC}_k|), \forall k=t_w,...,t_{end}\}
    \end{aligned}
    \label{eq:MAX}
\end{equation}{}

where, $L=t_{end}-t_w + 1$.\\  

In this paper, the MSE function is used as the objective function in equation \ref{eq:theta_argmin} and hence the CNN model is trained using backpropagation on the MSE loss function. The MSE loss which is also the L2 norm of the error vector is a widely used loss function in literature for training machine learning models (\cite{CHEMALI2018242}, \cite{8240689}, \cite{Li_2019}, \cite{9036949}). A Bayesian explanation that follows this preference is that the MSE loss corresponds to a KL divergence loss between probability densities of the estimated quantity and the true target variable, when the densities are assumed to be gaussian in nature. The gaussian assumption on probability densities is not a restrictive constraint as by the central limit theorem, presence of several different independent noises in the data implies that in limit the total noise in the data tends towards a gaussian distribution. This implies that in limit the data and some of its well behaved functions also tend to follow the gaussian distribution \cite{10.5555/1162264}. Moreover, the MSE or the L2 norm gives continuous gradients as it is differentiable at all points and gives a global minima only when the model exactly fits the data. It is because of these reasons that the mean squared error is used here as an objective function for training the proposed CNN model.
 
The MAE or the L1 norm and the MAX or the L-infinity norm of the error vector correspond to the average absolute first order deviation and the maximum absolute deviation of the model output from the true SoC values respectively. A low MAE and MAX implies that at almost all the sample points, the model output is closely following the actual target value. Moreover, unlike higher order loss functions, these first order metrics don't equate to unrealistically small values when the actual deviation is smaller than unity. Hence we have used them as validation metrics for testing the efficacy of the trained model on previously unseen data.
 
The CNN model is trained using a gradient based Adam optimizer and backpropagation is used to update its weights.
\vspace{-5pt}
\subsection{CNN for SoC Estimation}\label{subsec:CNN arch}

Convolutional Neural Networks are the state of the art deep learning algorithms for image processing. First introduced by Yann LeCun in \cite{726791}, the performance of CNNs is attributed to three major features embedded within its architecture - local receptive fields, shared weights and subsampling. The CNN proposed in this paper consists of one dimensional (1D) convolutional and pooling layers. Unlike in a standard 2D-CNN, where the kernels or filters stride across both the spatial dimensions of an image, i.e., from left to right and from top to bottom, kernels in 1D CNN layers stride only in one dimension, which is the temporal dimension in our case. Thus they are able to extract temporally relevant features. 

\begin{figure}[!ht]
\centering
\captionsetup{justification=centering}
\includegraphics[width=\columnwidth]{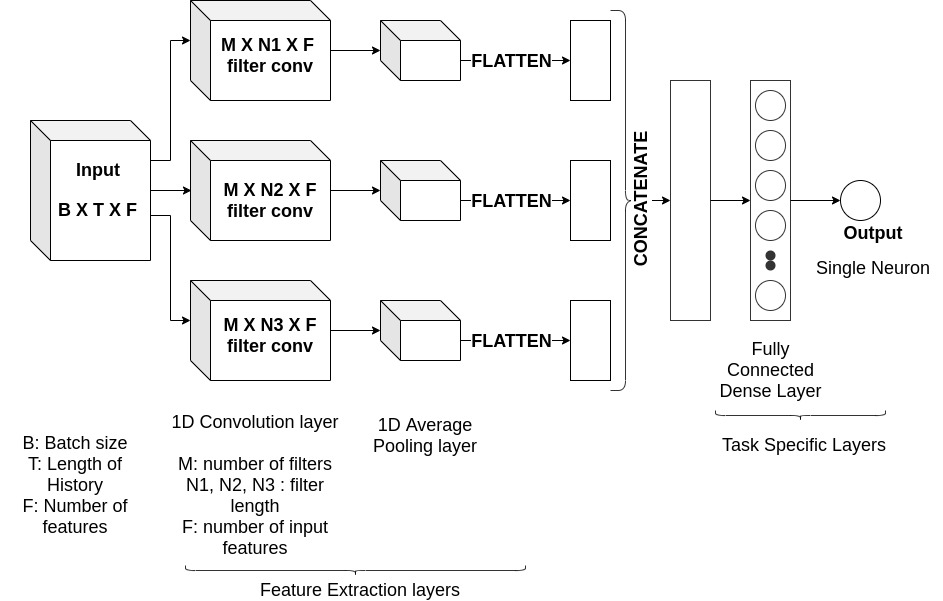}
\caption{Merge first architecture }
\label{fig:CDD}
\end{figure}

\begin{figure}[!ht]
  \centering
  \captionsetup{justification=centering}
  \includegraphics[width=\columnwidth]{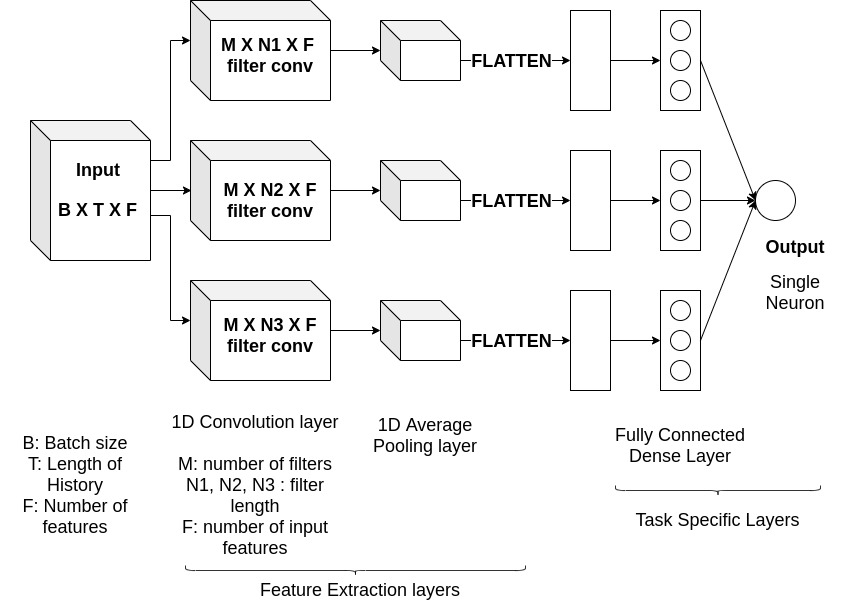}
  \caption{Dense first architecture}
  \label{fig:DCD}
\end{figure}

The proposed 1D CNN model takes the voltage, current and temperature values of a battery corresponding to timesteps $k-t_w+1$ to $k$ as inputs and predicts the SoC value at timestep $k$. Two basic CNN architectures, namely the dense first and the merge first architectures were developed for this application. As shown in figures \ref{fig:CDD} and \ref{fig:DCD}, the initial layers of the two architectures, consisting of convolutional and pooling layers are identical to each other. The first convolutional layer extracts three disjoint clusters of feature maps from the input. These clusters are formed by convoluting the input data sequence with three differently shaped filters or kernels of shapes $({t_w}/10) \times 3$ , $({t_w}/5) \times 3$ and $({t_w}/2) \times 3$ along the temporal axis. The first dimension of the filters corresponds to the number of datapoints along the time axis that are considered at once for convolution whereas the second dimension corresponds to the dimensionality of the feature space, which is 3, corresponding to battery voltage, current and temperature. The intention behind using filters of different shapes is to capture both long and short term temporal dependencies in the data. The subsequent convolutional and pooling layers follow suit and maintain the clustered feature map appearance where each cluster of feature maps in one layer is derived from its corresponding predecessor in the previous layer. The shape of the filters in the subsequent convolutional and pooling layers were kept the same for each of the three feature map clusters in a given layer.  

The difference in the architectures, however, is in terms of the positional arrangement of the final fully connected dense layers. In the merge first architecture, the feature maps extracted from the final pooling layer are concatenated after flattening and then given as input to the dense layer, whereas in the dense first architecture, the flattened feature maps are given as inputs to three disjoint fully connected dense layers and no concatenation operation is performed. The final neuron gives the SoC estimate in both the cases.

\begin{figure}[!ht]
  \centering
  \captionsetup{justification=centering}
  \includegraphics[width=\columnwidth]{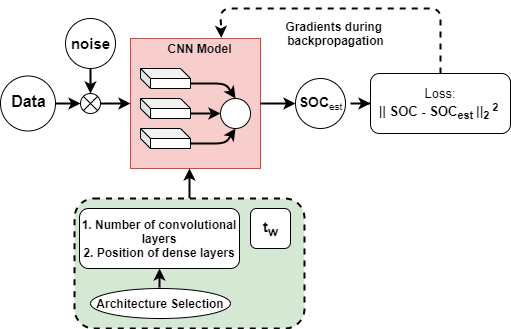}
  \caption{CNN Model Training Flowchart}
  \label{fig:CNN_training}
\end{figure}

The flowchart in figure \ref{fig:CNN_training} depicts the training procedure of the CNN based SoC Estimator.

{}

{}

{}

\begin{figure}[!ht]
  \centering
  \captionsetup{justification=centering}
  \includegraphics[width=\columnwidth]{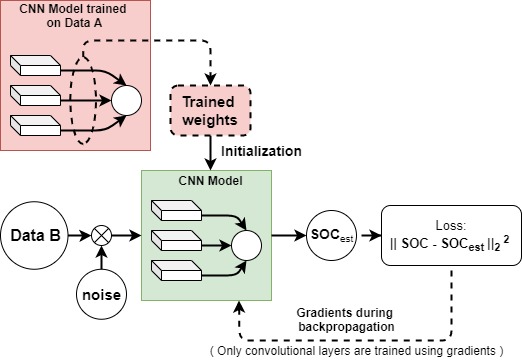}
  \caption{CNN Model training with Transfer Learning}
  \label{fig:CNN_training_TL}
\end{figure}{}

\subsection{Regularization in CNN model}\label{subsec:reg_intro}

Neural networks and deep learning algorithms, in general, can have a huge number of learnable parameters that makes them flexible and accurate model estimators. But it also renders them vulnerable towards noise as they can easily learn the noise in data along with the underlying pattern in it, which leads to overfitting and results in the model performing badly on previously unseen data and leads to high variance in predictions. In order to prevent this, certain restrictive conditions are enforced on the learnable weights of these models which are known as regularization techniques. Regularization helps in decreasing the variance of predictions, increasing the generalizability of the model and making the model robust to noise \cite{10.5555/1162264}. \\
CNNs themselves come with innate regularizations embedded in their architecture in the form of local receptive fields, shared weights within filters, subsampling or pooling operations, etc. Along with them, some external regularization techniques have also been used in the proposed CNN model which are given below:
\begin{enumerate}
\item L2 regularization in final layer: An L2 norm of the final layer weight vector is added to the training loss, thus forcing its weights to stay as low as possible. A regularization coefficient of $1e-4$ is used in this paper.

\item Dropout in convolutional layers: The weights of the convolutional layers where dropout is implemented are dropped with a probability of $20\%$.

\end{enumerate}

\subsection{Knowledge Transfer across datasets}\label{subsec:TL_intro}

A typical machine learning (ML) problem can be characterized by a domain and a task. The domain comprises of the feature space $\mathcal{X}$ and the marginal feature distribution $\mathbb{P}(X)$ while the task comprises of the label space $\mathcal{Y}$ and the posterior of the labels conditioned on features, i.e., $\mathbb{P}(Y|X)$. It has been observed that an ML algorithm, especially deep learning algorithm trained on a specific task and domain performs poorly when either of them changes during testing.

A transfer learning algorithm formally identifies a source domain $\mathfrak{D_s}$ and a corresponding task $\mathfrak{T_s}$ and a target domain $\mathfrak{D_t}$ and a corresponding task $\mathfrak{T_t}$ and aims at transferring knowledge across them.
Depending on what changes in the domain or the task when we shift from a source problem to a new target problem, transfer learning can be classified as inductive, transductive and unsupervised \cite{5288526}. The SoC estimation problem across different battery types falls under the transductive class of transfer learning algorithms that are used when the source and target domains are different but related. More specifically, we assume that the feature spaces are the same but the marginal feature distribution are different, i.e.,$X_\mathfrak{s}$ $\neq$  $X_\mathfrak{t}$  $\Rightarrow$ $\mathbb{P}_\mathfrak{s}(X)$ $\neq$ $\mathbb{P}_\mathfrak{t}(X)$. The corresponding tasks are the same, i.e., $\mathfrak{T_s}=\mathfrak{T_t}$.

In order to establish the benefits of using transfer learning empirically, we use two publicly available battery datasets in this paper- Dataset A: Panasonic 18650 PF battery dataset , Dataset B: LG 18650 dataset. Details of the datasets are given in section \ref{sec:data desc}. 

Considering the fact that labelled data is available in case of both the source and target problems, a simple weight sharing algorithm is used for implementing transfer learning in this paper. A CNN model, model A is first trained on Dataset A (source data). Then the final weights of model A is used as the initial set of weights for training an identical model, model B, on Dataset B (target data). The fully connected dense layer weights of model B are frozen after initialization and only convolutional layers are updated during training with Dataset B.\\ 
Such intelligent initializations allow model B to converge quickly to the optimal point and with much less amount data from Dataset B than is required when randomly initialized. The intuition behind training just the initial layers of the CNN model B arises from the fact that the initial layers perform the function of feature selection and since the source and target domains are different, the features selected for the task corresponding to the target domain need to be different than those used for the source task. The flowchart in figure \ref{fig:CNN_training_TL} depicts the training procedure of a transfer learning model.

\section{Dataset description}\label{sec:data desc}
Two publicly available LiB datasets have been used for experimentation in this paper, the detailed specifications of which can be found in \cite{k18} and \cite{kvns20}. The drive cycle data in both these datasets is originally sampled at a frequency of 10 Hz. However, for each of the two battery types, keeping in mind the practical limitations of a Battery Management System, along with the original LiB dataset sampled at 10Hz, two new datasets were formed by downsampling the original dataset to 1 Hz and 0.1 Hz. All further analysis in this paper has been carried out on datasets of all the three available frequencies.

\subsection{Dataset A – Panasonic 18650 PF battery dataset}\label{subsec:dataA}
\begin{figure}[!ht]
    \centering
    \captionsetup{justification=centering}
    \includegraphics[keepaspectratio, width=\linewidth]{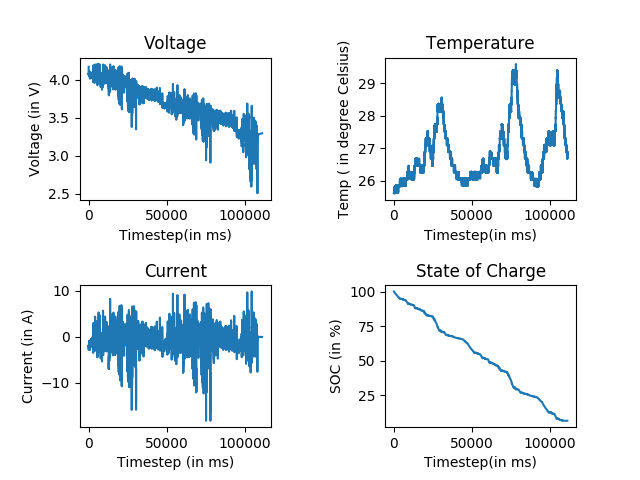}
    \caption{Cycle 2 drive cycle data at 25\textdegree C from the Panasonic Dataset}
    \label{fig:Pan_Cycle2}
\end{figure}

This dataset was prepared in the University of Wisconsin Madison \cite{k18}. The battery cell consists of a Lithium Nickel Cobalt Aluminimum Oxide ($LiNiCoAlO_2$ or NCA) chemistry. The data collected in this dataset corresponds to tests conducted at different ambient temperatures ranging from $-20$ \textdegree C to $25$ \textdegree C. For each test the battery cell was first fully charged and then a power drive cycle profile was executed on it. The drive cycles correspond to a Ford F150 truck and were drawn from the battery cell during discharge till the cell voltage reached its cut off value of 2.5V. The four standard drive cycles that the battery was put to test with were –Urban Dynamometer Driving Schedule (UDDS) , Highway Fuel Economy Driving Schedule (HWFET) , Los
Angeles 92 (LA92) and Supplemental Federal Test Procedure Driving
Schedule (US06). Some additional drive cycles were manually created for obtaining additional dynamics. These cycles, namely, Cycle 1, Cycle 2, Cycle 3, Cycle 4, and Neural Network (NN), consist of a random mix of UDDS, HWFET, US06 and LA92. Thus data corresponding to a total of 9 drive cycles are available. The charging dynamics are always the same. Data corresponding to Cycle 2 at an ambient temperature of $25$ \textdegree C is plotted in figure \ref{fig:Pan_Cycle2}. The state of charge as can be seen has a highly non linear characteristic.
{}

\begin{figure}[!ht]
    \centering
    \captionsetup{justification=centering}
    \includegraphics[keepaspectratio, width=\linewidth]{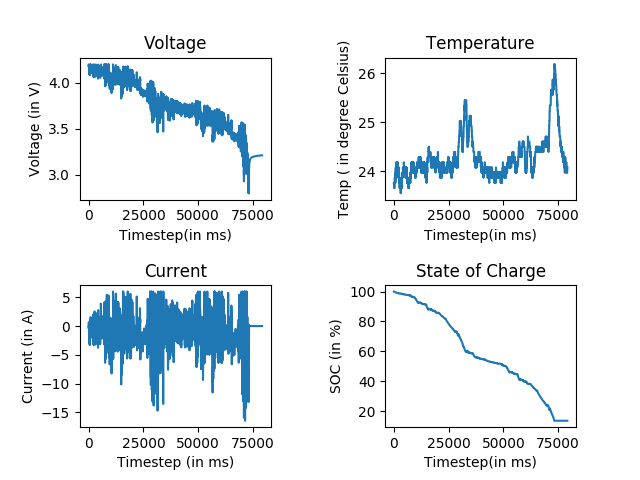}
    \caption{Mixed 2 drive cycle data at 25\textdegree C from the LG 18650 HG2 Dataset}
    \label{fig:LG_Mixed2}
\end{figure}{}

\subsection{Dataset B – LG 18650 HG2 battery dataset}\label{subsec:dataB}
This dataset was prepared in the McMaster University, Ontario, Canada \cite{kvns20}. A $3Ah$ LG HG2 cell, with a Lithium Nickel Manganese Cobalt oxide ($Li[NiMnCo]O_2$ or NMC) chemistry, was put under a series of tests under six different ambient temperatures ranging from $-20$ \textdegree C to $40$ \textdegree C. The tests conducted were similar to that on Dataset A. Four standard drive cycles were drawn from the battery – UDDS, LA92, HWFET and US06. Apart form these eight additional cycles were artificially created, namely, Mixed 1, Mixed 2, Mixed 3, Mixed 4, Mixed 5, Mixed 6, Mixed 7 and Mixed 8. These cycles are a random mix of the four standard drive cycles. The battery was fully charged before each test and was discharged through these drive cycle profiles till $95\%$ of the $1$C discharge capacity at the respective temperature was drawn from the cell. Figure \ref{fig:LG_Mixed2} consists of the visual depiction of data corresponding to relevant battery parameters for Mixed 2 cycle at $25$
\textdegree C. 
%\vspace{-5pt}
%\vspace{-5pt}
\subsection{Additive Noise in Data}\label{subsec:noise}
In order to test the robustness of the proposed algorithm against noise in data, two different types of noises were added to the data features post normalization. The noises are independent across each data dimension. They are detailed below:
\begin{enumerate}
    \item Noise A: 
        $\epsilon \sim \mathcal{N}(0, 0.01)$
    \item Noise B:\\
   \\
    ${z_0},{z_1} \sim \mathcal{U}(1,10)$\\
    $x \sim \mathcal{N}({z_0},{z_1})$\\
    $\eta \sim \mathcal{U}(1, 5)$\\
    $ a = \sin(a\eta+\tanh(x))$\\
    $\epsilon = 1/ (1 + \exp(0.3a))$\\
\end{enumerate}

Here $\mathcal{N}$ refers to a gaussian distribution and $\mathcal{U}$ refers to a uniform distribution.
%\begin{figure}[!ht]
%    \centering
%    \captionsetup{justification=centering}
%    \includegraphics[keepaspectratio, width=\linewidth]{images/Mixed2_LG_25.png}
%    \caption{Mixed 2 drive cycle data at 25\textdegree C from the LG 18650 HG2 Dataset}
%    \label{fig:LG_Mixed2}
%\end{figure}{}
%\vspace{-5pt}
%\vspace{-5pt}
\section{Experimental Setup}\label{sec:exp setup}
The experiments conducted for this paper can be broadly categorized into three subsequent sections as elaborated below.

\subsection{Architecture Selection}\label{subsec:arch sel}
These are the first set of experiments conducted in this paper. The aim of these experiments is to identify the best values for two important hyper-parameters, namely the number of convolutional layers and the position of the fully connected dense layers, to be used to define the 1D CNN architecture for the subsequent experiments. To consider the effect of varying the position and organization of the fully connected dense layers, the two architectures depicted in figures 1 and 2 are tested. For each of these two basic architectures the number of convolutional layers are varied between 1 and 2. Hence a total of four CNN architectures are tested in these set of experiments.For these experiments, only the Panasonic battery dataset sampled at 10 Hz was used and $t_w$ was fixed at 500. \\ 
Seven drive cycles, namely, Cycle 1 to 4 , UDDS and LA92 from Dataset A are considered for training a particular architecture of the model at different ambient temperatures individually. For each architecture under consideration, a model was independently trained on three different datasets, each consisting of the aforementioned seven drive cycles recorded at one of the three considered ambient temperatures - 0°C, 10°C and 25°C. After training, each architecture is evaluated on the remaining two drive cycles – US06 and HWFET - recorded at ambient temperatures corresponding to the dataset that they were trained with and the MAE and MAX errors of estimation were recorded. The model architecture which gives the least error in estimation under all the three ambient temperatures considered is used for further experiments.
The entire architecture of the finally selected model is given in table \ref{tab:table3}.

\begin{table}[!ht]
\begin{center}
\caption{ Architecture of the best performing model obtained from the architecture selection experiment
}
\label{tab:table3}
\resizebox{\columnwidth}{!}{
\renewcommand{\arraystretch}{1.5}
\begin{tabular}{|c|c|}
\hline
\textbf{Hyperparameters}                                                                                                        & \textbf{Values}                                                                       \\ \hline
Number of convolutional layers                                                                                                  & 2                                                                                     \\ \hline
Architecture Type                                                                                                               & Dense First                                                                           \\ \hline
\begin{tabular}[c]{@{}c@{}}Number of neurons in each individual \\ dense layer cluster in the \\ penultimate layer\end{tabular} & 64                                                                                    \\ \hline
Shapes of filters in 1st convolutional layer                                                                                    & $t_w/10$, $t_w/5$ and $t_w/2$                                                                        \\ \hline
\begin{tabular}[c]{@{}c@{}}Number of filters of each type \\ in the 1st conv layer\end{tabular}                                 & 16                                                                                    \\ \hline
Shapes of filters in 2nd  convolutional layer                                                                                   & $t_w/5$, $t_w/5$ and $t_w/5$                                                                         \\ \hline
\begin{tabular}[c]{@{}c@{}}Number of filters of each type \\ in the 2nd conv layer\end{tabular}                                 & 8                                                                                     \\ \hline
Number of pooling layers                                                                                                        & 1                                                                                     \\ \hline
Position of pooling layer                                                                                                       & After the 2nd convolutional layer                                                     \\ \hline
Type of pooling                                                                                                                 & Average Pooling                                                                       \\ \hline
Shapes of pooling filters                                                                                                       & 10, 10 and 10                                                                         \\ \hline
Activation used in other than final layer                                                                                       & Leaky ReLu                                                                            \\ \hline
Activation used in final layer                                                                                                  & ReLu                                                                                  \\ \hline
Regularizer                                                                                                                     & \begin{tabular}[c]{@{}c@{}}L2 with coeff of 1e-4 \\ (in the final layer)\end{tabular} \\ \hline
Dropout probability                                                                                                             & 0.2                                                                                  \\ \hline
\end{tabular}}
\end{center}
\end{table}
%\vspace{-5pt}
\begin{figure}[!ht]
  \centering
  \captionsetup{justification=centering}
  \includegraphics[width=\columnwidth]{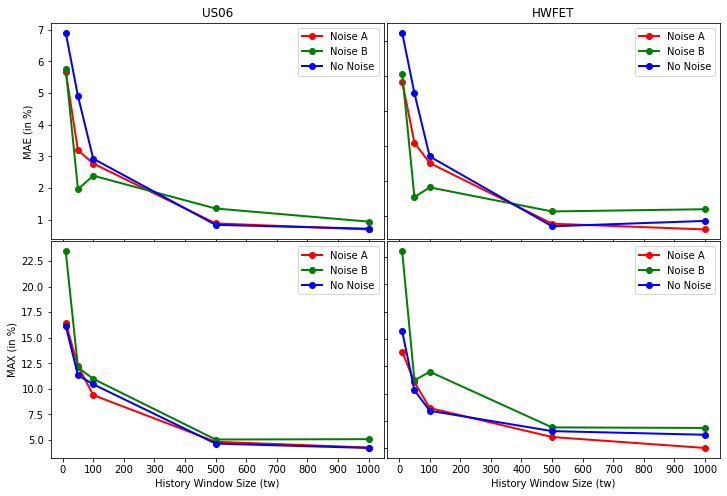}
  \caption{Variations in MAE and MAX values obtained on US06 and HWFET drive cycles of the Panasonic battery dataset, sampled at 10 Hz, depending on the type of additive noise in data and history size $t_w$} 
  \label{fig:Pan_loss_var_original}
\end{figure}{}

\begin{figure}[!ht]
  \centering
  \captionsetup{justification=centering}
  \includegraphics[width=\columnwidth]{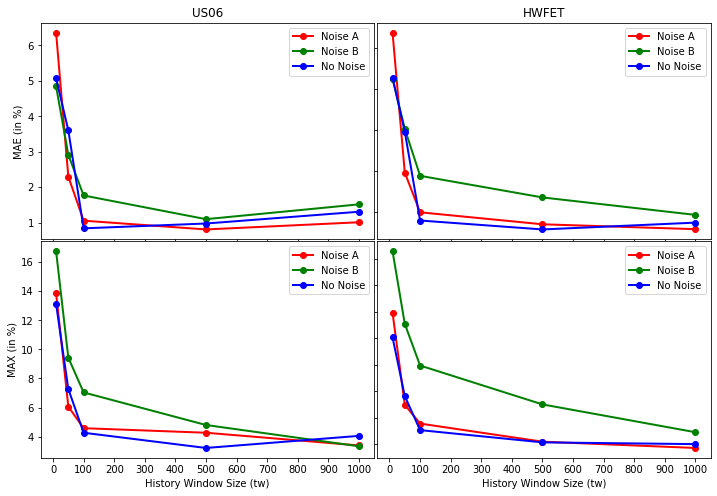}
  \caption{Variations in MAE and MAX values obtained on US06 and HWFET drive cycles of the Panasonic battery dataset, sampled at 1 Hz, depending on the type of additive noise in data and history size $t_w$ }
  \label{fig:Pan_loss_var_down_by_10}
\end{figure}{}

\begin{figure}[!ht]
  \centering
  \captionsetup{justification=centering}
  \includegraphics[width=\columnwidth]{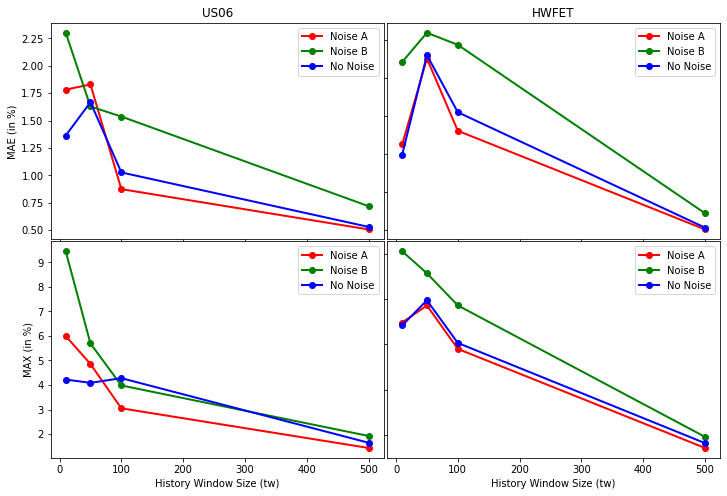}
  \caption{Variations in MAE and MAX values obtained on US06 and HWFET drive cycles of the Panasonic battery dataset, sampled at 0.1 Hz, depending on the type of additive noise in data and history size $t_w$} 
  \label{fig:Pan_loss_var_down_by_100}
\end{figure}{}

\subsection{Model training and evaluation}\label{subsec:mod train}
With the best performing basic architecture obtained from the previous experiment, the CNN model is trained on a larger battery dataset formed by combining data corresponding to all the available ambient temperatures. The effect of the window size, $t_w$, is studied by varying it between five values given by \{10,50,100,500,1000\}. In order to test the effect of data timestep on SoC estimation, the existing data from both the LG and Panasonic batteries which were recorded at a timestep of 0.01s, i.e. with a frequency of 10Hz, were downsampled by factors of 10 and 100 as mentioned previously in Section III. That is, for each battery, three different datasets, sampled at 10Hz, 1Hz and 0.1 Hz were formed. Moreover, in addition to using the dataset directly as available online, noise A and noise B as specified in Section \ref{subsec:noise} were added to the dataset to test the efficacy of the algorithm under noisy data conditions.\\
For the Panasonic dataset, the training data consisted of Cycles 1 to 4 and UDDS corresponding to all the available ambient temperatures, i.e., -20°C, -10°C, 0°C , 10°C and 25°C. The validation data consisted of the LA92 drive cycle corresponding to these temperatures. The model was tested on US06 and HWFET cycles corresponding to these temperatures.\\
For the LG dataset, the training data consisted of cycles Mixed 1-8 and UDDS, validation dataset consisted of LA92 and the test data consisted of US06 and HWFET cycles at the ambient temperatures given by -20°C, -10°C, 0°C , 10°C and 25°C.\\
The MAE and MAX errors were recorded and the final trained weights of the model were stored. Training was carried out for 100 epochs in a mini-batch fashion with a batch size of 128. The training was stopped when validation error didn’t decrease for 10 consecutive epochs.
\begin{figure}[!ht]
  \centering
  \captionsetup{justification=centering}
  \includegraphics[width=\columnwidth]{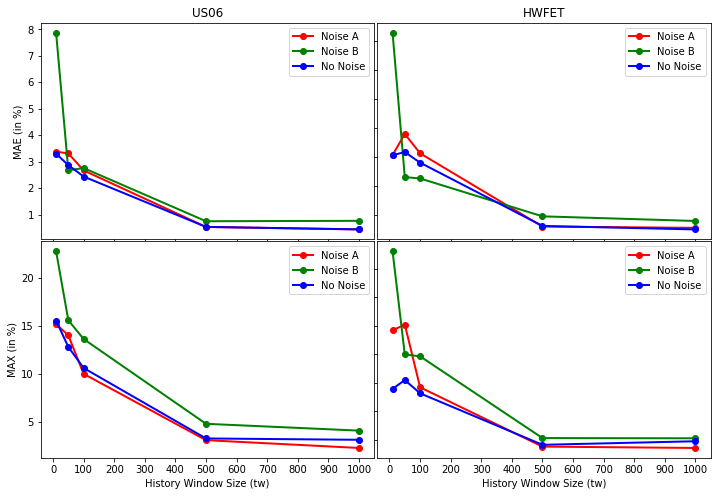}
  \caption{Variations in MAE and MAX values obtained on US06 and HWFET drive cycles of the LG battery dataset, sampled at 10 Hz, depending on the type of additive noise in data and history size $t_w$} 
  \label{fig:LG_loss_var_original}
\end{figure}{}

\begin{figure}[!ht]
  \centering
  \captionsetup{justification=centering}
  \includegraphics[width=\columnwidth]{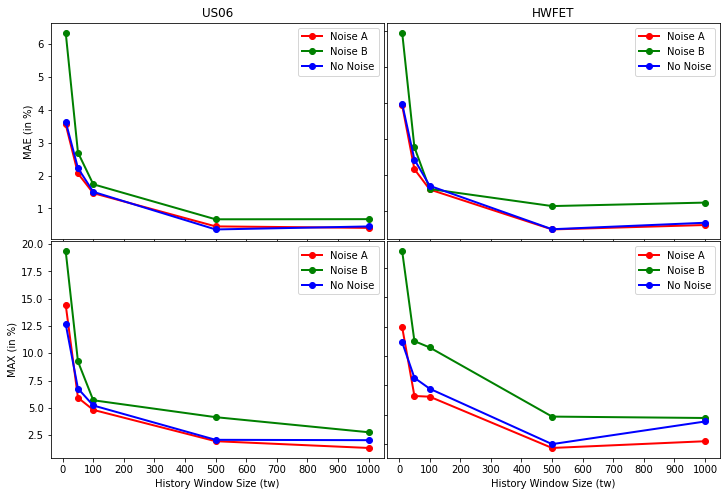}
  \caption{Variations in MAE and MAX values obtained on US06 and HWFET drive cycles of the LG battery dataset, sampled at 1 Hz, depending on the type of additive noise in data and history size $t_w$} 
  \label{fig:LG_loss_var_down_by_10}
\end{figure}{}

\begin{figure}[!ht]
  \centering
  \captionsetup{justification=centering}
  \includegraphics[width=\columnwidth]{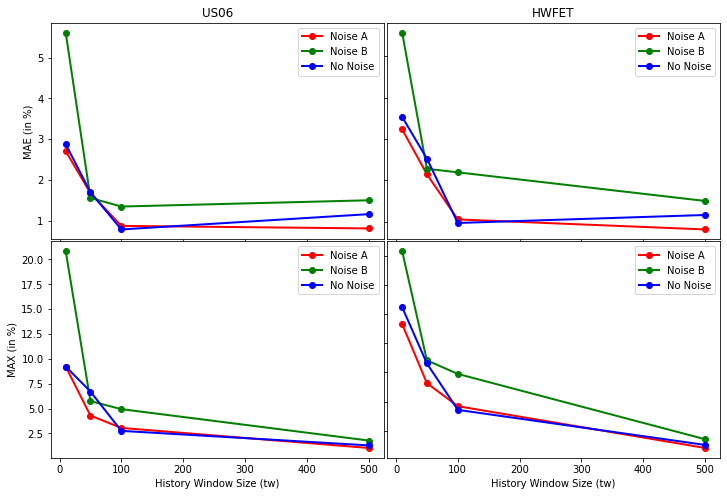}
  \caption {Variations in MAE and MAX values obtained on US06 and HWFET drive cycles of the LG battery dataset, sampled at 0.1 Hz, depending on the type of additive noise in data and history size $t_w$} 
  \label{fig:LG_loss_var_down_by_100}
\end{figure}{}

\subsection{Transfer learning on different dataset}\label{subsec:TL_exp}
In these set of experiments, the weights of a model trained on the Panasonic dataset (source data) is used to initialize an identical model to be trained on the LG dataset (target data) as described in section \ref{subsec:TL_intro}. As the target data, LG datasets sampled at 10Hz frequency and 1Hz frequency were used. The effects of noises A and B, corresponding to those described in section \ref{subsec:noise}, on the data is considered. In each of the scenarios considered in this experiment, the CNN model to be trained on a particularly conditioned LG dataset is initialized with the trained weights of an identical CNN model trained on a similarly conditioned Panasonic dataset. For example, a CNN model with the history window size, $t_w$ =500 to be trained on the LG dataset, sampled at 1 Hz and with an added noise A will be initialized by the weights of an identical CNN model with $t_w$=500 that was trained on the Panasonic dataset, sampled at 1 Hz and with added noise A. The training of this newly initialized model on the LG dataset is carried out according to the procedure described in section \ref{subsec:TL_intro} and depicted in figure \ref{fig:CNN_training_TL}.\\
The drive cycles and the corresponding ambient temperatures used for the training, test and validation datasets in this experiment is similar to what has been used for the LG dataset in the previous experiments. Additionally, two scenarios are looked into while deciding how much of training data will be used for training the CNN model initialized using transfer learning. Depending on how much training data will be used, the two scenarios are explained below:
\begin{enumerate}
    \item With full LG data:\\
The entire LG training dataset, consisting of 45 drive cycles, corresponding to the different ambient temperatures as specified in section \ref{subsec:mod train} is used to train the transfer learning CNN model. The values of $t_w$ are varied between \{1000,500,100\}. This is because the CNN model has been observed to perform badly for lesser values of $t_w$.
    \item With $40\%$ LG data: \\
In this case, each of the 45 drive cycles are kept in the training data with a probability of 0.4. The value of $t_w$ is fixed for this experiment at 500. The effects of noise has been studied.
\end{enumerate}

Training was carried out for a maximum of 100 epochs in the mini-batch fashion with a batch size of 128 and early stopping was implemented based on validation error.

\begin{figure}[!ht]
  \centering
  \captionsetup{justification=centering}
  \includegraphics[width=\columnwidth]{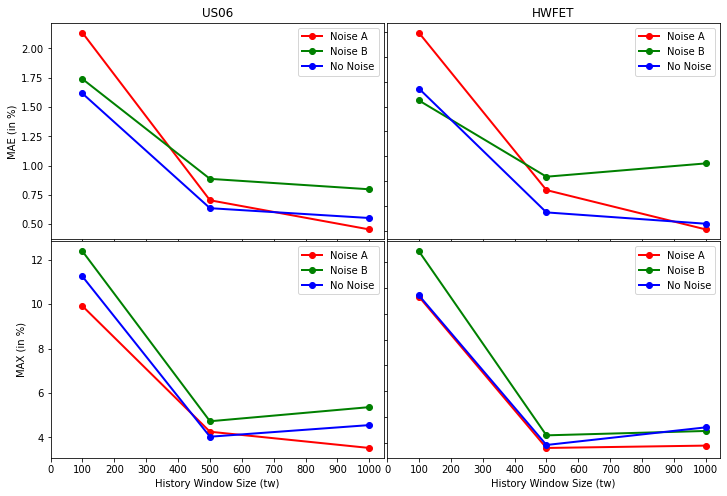}
  \caption{Transfer Learning Results: Variations in MAE and MAX values obtained on US06 and HWFET drive cycles of the LG battery dataset after training the CNN model using transfer learning with the entire LG training data, sampled at 10 Hz. The variations are shown with respect to the type of additive noise in data and history size $t_w$} 
  \label{fig:LG_TL_loss_original}
\end{figure}{}

\begin{figure}[!ht]
  \centering
  \captionsetup{justification=centering}
  \includegraphics[width=\columnwidth]{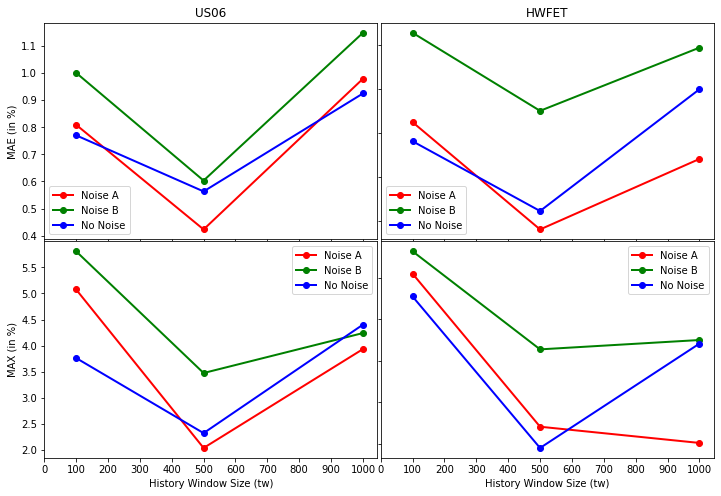}
  \caption{Transfer Learning Results: Variations in MAE and MAX values obtained on US06 and HWFET drive cycles of the LG battery dataset after training the CNN model using transfer learning with the entire LG training data, sampled at 1 Hz. The variations are shown with respect to the type of additive noise in data and history size $t_w$} 
  \label{fig:LG_TL_loss_down_by_10}
\end{figure}{}

\begin{figure}[!ht]
  \centering
  \captionsetup{justification=centering}
  \includegraphics[width=\columnwidth]{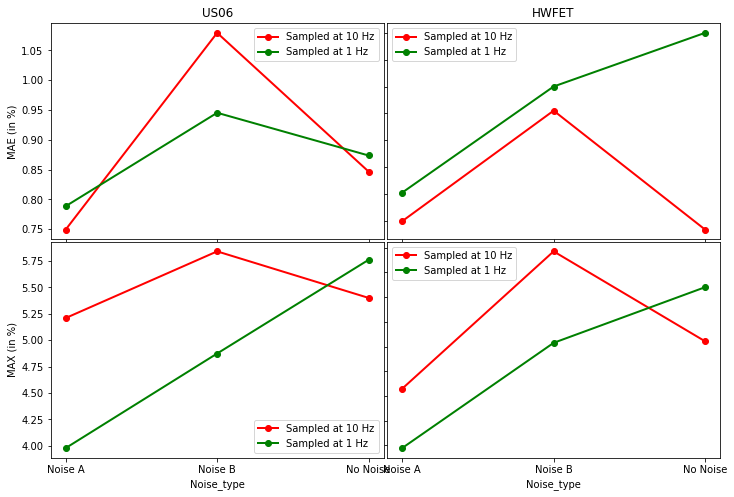}
  \caption{Transfer Learning Results: Variations in MAE and MAX values obtained on US06 and HWFET drive cycles of the LG battery dataset after training the CNN model using transfer learning with only 40\% of the training cycles. The variations are shown with respect to the type of additive noise in data and data sampling frequency. The $t_w$ is fixed at 500.}
  \label{fig:LG_loss_TL_ig_60}
\end{figure}{}

\section{Results and Discussions}\label{sec:results}

The following section consists of the results of the various experiments described in the previous section and subsequent inferences derived from the observed results.
\subsection{Architecture Selection}\label{subsecexp:arch sel}

The final results comprising of MAE and MAX errors corresponding to each of the tested architectures and measured on the test datasets at each temperature are given in Table \ref{tab:table4}.

\begin{table*}[!ht]
\begin{center}
\caption{ Estimation error for the architecture selection experiment. The errors under each drive cycle is in the form MAE ($\%$) / MAX($\%$)}
\label{tab:table4}
\resizebox{\textwidth}{!}{
\renewcommand{\arraystretch}{1.2}
\begin{tabular}{|c|c|c|c|c|c|c|c|}
\hline
\multirow{3}{*}{\textbf{Architecture type}} & \multirow{3}{*}{\textbf{Number of convolutional layers}} & \multicolumn{6}{c|}{\textbf{Temperature 			(in degree Celsius)}}                                                \\ \cline{3-8} 
                                            &                                                          & \multicolumn{2}{c|}{\textbf{0}} & \multicolumn{2}{c|}{\textbf{10}} & \multicolumn{2}{c|}{\textbf{25}} \\ \cline{3-8} 
                                            &                                                          & \textbf{US06}  & \textbf{HWFET} & \textbf{US06}  & \textbf{HWFET}  & \textbf{US06}  & \textbf{HWFET}  \\ \hline
Merge first                                 & 1                                                        & 0.98/4.27      & 1.33/3.61      & 0.76/3.41      & 1.18/5.23       & 0.89/3.87      & 0.99/4.26       \\ %\hline
Merge first                                 & 2                                                        & 1.18/7.55      & 1.59/3.98      & 0.64/4.96      & 1.16/5.23       & 0.75/2.9       & 0.89/4.21       \\ %\hline
Dense first                                 & 1                                                        & 0.91/5.12      & 1.34/3.93      & 0.82/6.1       & 1.26/4.66       & 0.96/3.68      & 1.123/4.42      \\ %\hline
Dense first                                 & 2                                                        & 1.25/6.5       & 1.62/3.66      & 0.85/3.5       & 1.13/5.09       & 0.58/2.78      & 0.77/4.33       \\ \hline
\end{tabular}}
\end{center}
\end{table*}

\begin{table*}[!ht]
\begin{center}
\caption{SoC estimation error values obtained on the test datasets derived from Panasonic Battery dataset for the $2C-DF$ architecture. The errors under each drive cycle is in the form MAE ($\%$) / MAX($\%$). $f_s$ stands for sampling frequency of data.}
\label{tab:tablePan}
\resizebox{\textwidth}{!}{
\renewcommand{\arraystretch}{1.5}
\begin{tabular}{|c|c|c|c|c|c|c|c|c|c|c|}
\hline
\multirow{2}{*}{\textbf{$f_s$ (Hz)}} & \multirow{2}{*}{\textbf{Cycle}} & \multicolumn{3}{c|}{\textbf{$t_w$ = 1000}}                                                                & \multicolumn{3}{c|}{\textbf{$t_w$ = 500}}                                                                           & \multicolumn{3}{c|}{\textbf{$t_w$ = 100}}                                                                 \\ \cline{3-11} 
                                     &                                 & \multicolumn{1}{c|}{\textbf{Noise A}} & \multicolumn{1}{c|}{\textbf{Noise B}} & \textbf{No Noise} & \multicolumn{1}{c|}{\textbf{Noise A}} & \multicolumn{1}{c|}{\textbf{Noise B}} & \textbf{No Noise}           & \multicolumn{1}{c|}{\textbf{Noise A}} & \multicolumn{1}{c|}{\textbf{Noise B}} & \textbf{No Noise} \\ \hline
\multirow{2}{*}{\textbf{10}}         & \textbf{US06}                   & 0.68/ 4.26                            & 0.93/ 5.08                            & 0.71/ 4.22        & 0.87/ 4.82                            & 1.34/ 5.05                            & 0.83/ 4.65                  & 2.76/ 9.38                            & 2.38/ 10.76                           & 2.92/ 10.44       \\ \cline{2-11} 
                                     & \textbf{HWFET}                  & 0.62/ 2.53                            & 1.20/ 4.37                            & 0.86/ 3.74        & 0.78/ 3.53                            & 1.13/ 4.42                            & 0.71/ 4.08                  & 2.51/ 6.19                            & 1.82/ 9.51                            & 2.70/ 5.94        \\ \hline
\multirow{2}{*}{\textbf{1}}          & \textbf{US06}                   & 1.01/ 3.40                            & 1.51/ 3.34                            & 1.3/ 4.00         & 0.8/ 4.27                             & 1.09/ 4.80                            & 0.97/ 3.22                  & 1.05/ 4.58                            & 1.76/ 7.04                            & 0.83/ 4.27        \\ \cline{2-11} 
                                     & \textbf{HWFET}                  & 0.57/ 1.70                            & 0.92/ 2.89                            & 0.73/ 2.00        & 0.69/ 2.18                            & 1.34/ 5.00                            & 0.57/ 2.13                  & 0.98/ 3.55                            & 1.87/ 7.93                            & 0.78/ 3.05        \\ \hline
\multirow{2}{*}{\textbf{0.1}}        & \textbf{US06}                   & -                                     & -                                     & -                 & \multirow{2}{*}{0.50/ 1.43}           & \multirow{2}{*}{0.71/ 1.92}           & \multirow{2}{*}{0.53/ 1.64} & 0.87/ 3.05                            & 1.53/ 3.98                            & 1.02/ 4.27        \\ \cline{2-5} \cline{9-11} 
                                     & \textbf{HWFET}                  & -                                     & -                                     & -                 &                                       &                                       &                             & 1.80/ 5.80                            & 2.93/ 7.72                            & 2.04/ 6.06        \\ \hline
\end{tabular}
}
\end{center}
\end{table*}

\begin{table*}[!ht]
\begin{center}
\caption{ SoC estimation error values obtained on the test datasets derived from LG Battery dataset for the $2C-DF$ architecture. The errors under each drive cycle is in the form MAE ($\%$) / MAX($\%$). $f_s$ stands for sampling frequency of data.}
\label{tab:tableLG}
\resizebox{\textwidth}{!}{
\renewcommand{\arraystretch}{1.5}

\begin{tabular}{|c|c|c|c|c|c|c|c|c|c|c|}
\hline
\multirow{2}{*}{\textbf{$f_s$ (Hz)}} & \multirow{2}{*}{\textbf{Cycle}} & \multicolumn{3}{c|}{\textbf{$t_w$ = 1000}}                      & \multicolumn{3}{c|}{\textbf{$t_w$ = 500}}                                                       & \multicolumn{3}{c|}{\textbf{$t_w$ = 100}}                       \\ \cline{3-11} 
                                     &                                 & \textbf{Noise A} & \textbf{Noise B} & \textbf{No Noise} & \textbf{Noise A}            & \textbf{Noise B}            & \textbf{No Noise}           & \textbf{Noise A} & \textbf{Noise B} & \textbf{No Noise} \\ \hline
\multirow{2}{*}{\textbf{10}}         & \textbf{US06}                   & 0.43/ 2.30       & 0.76/ 4.11       & 0.44/ 3.15        & 0.53/ 3.11                  & 0.74/ 4.82                  & 0.53/ 3.29                  & 2.66/ 10.03      & 2.75/ 13.66      & 2.43/ 10.63       \\ \cline{2-11} 
                                     & \textbf{HWFET}                  & 0.54/ 4.28       & 0.78/ 5.13       & 0.48/ 4.87        & 0.58/ 4.40                  & 0.94/ 5.15                  & 0.61/ 4.55                  & 3.12/ 9.60       & 2.25/ 12.33      & 2.79/ 9.08        \\ \hline
\multirow{2}{*}{\textbf{1}}          & \textbf{US06}                   & 0.41/ 1.33       & 0.67/ 2.77       & 0.45/ 2.04        & 0.45/ 1.95                  & 0.67/ 4.14                  & 0.36/ 2.07                  & 1.46/ 4.81       & 1.73/ 5.69       & 1.50/ 5.21        \\ \cline{2-11} 
                                     & \textbf{HWFET}                  & 0.61/ 2.75       & 1.24/ 4.73       & 0.68/ 4.43        & 0.49/ 2.18                  & 1.14/ 4.85                  & 0.5/ 2.50                   & 1.60/ 6.54       & 1.61/ 10.73      & 1.70/ 7.22        \\ \hline
\multirow{2}{*}{\textbf{0.1}}        & \textbf{US06}                   & -                & -                & -                 & \multirow{2}{*}{0.81/ 1.03} & \multirow{2}{*}{1.50/ 1.78} & \multirow{2}{*}{1.16/ 1.29} & 0.87/ 3.04       & 1.34/ 4.93       & 0.78/ 2.75        \\ \cline{2-5} \cline{9-11} 
                                     & \textbf{HWFET}                  & -                & -                & -                 &                             &                             &                             & 1.05/ 4.60       & 2.19/ 7.36       & 0.96/ 4.29        \\ \hline
\end{tabular}
}
\end{center}
\end{table*}

\begin{table*}[!ht]
\begin{center}
\caption{ Transfer Learning Results: SoC estimation error values obtained on the test datasets derived from LG Battery dataset for the $2C-DF$ architecture with transfer learning. Results for both the cases: one where the entire LG training data is used and the other where 60\% of the training data is stochastically eliminated are given. The errors under each drive cycle is in the form MAE ($\%$) / MAX($\%$). $f_s$ stands for sampling frequency of data.}
\label{tab:tableLG_TL}
\resizebox{\textwidth}{!}{
\renewcommand{\arraystretch}{2.0}
\begin{tabular}{|c|c|c|c|c|c|c|c|c|c|c|c|c|c|}
\hline
\multirow{2}{*}{\textbf{$f_s$ (Hz)}} & \multirow{2}{*}{\textbf{Cycle}} & \multicolumn{9}{c|}{\textbf{Full data}}                                                                                                                                     & \multicolumn{3}{c|}{\textbf{60\% data ignored}}         \\ \cline{3-14} 
                                     &                                 & \multicolumn{3}{c|}{\textbf{$t_w$ = 1000}}                      & \multicolumn{3}{c|}{\textbf{$t_w$ = 500}}                       & \multicolumn{3}{c|}{\textbf{$t_w$ = 100}}                       & \multicolumn{3}{c|}{\textbf{$t_w$ = 500}}                       \\ \hline
\multicolumn{2}{|c|}{\textbf{}}                                        & \textbf{Noise A} & \textbf{Noise B} & \textbf{No Noise} & \textbf{Noise A} & \textbf{Noise B} & \textbf{No Noise} & \textbf{Noise A} & \textbf{Noise B} & \textbf{No Noise} & \textbf{Noise A} & \textbf{Noise B} & \textbf{No Noise} \\ \hline
\multirow{2}{*}{\textbf{10}}         & \textbf{US06}                   & 0.45/ 3.52       & 0.79/ 5.36       & 0.55/ 4.55        & 0.70/ 4.25       & 0.88/ 4.73       & 0.63/ 4.03        & 2.13/ 9.91       & 1.73/ 12.37      & 1.61/ 11.24       & 0.74/ 5.20       & 1.07/ 5.84       & 0.84/ 5.40        \\ \cline{2-14} 
                                     & \textbf{HWFET}                  & 0.51/ 4.88       & 1.17/ 5.45       & 0.57/ 5.59        & 0.90/ 4.79       & 1.04/ 5.28       & 0.68/ 4.90        & 2.49/ 10.66      & 1.81/ 12.42      & 1.93/ 10.71       & 0.79/ 5.05       & 1.21/ 6.17       & 0.76/ 5.44        \\ \hline
\multirow{2}{*}{\textbf{1}}          & \textbf{US06}                   & 0.97/ 3.93       & 1.14/ 4.24       & 0.92/ 4.40        & 0.42/ 2.03       & 0.6/ 3.47        & 0.56/ 2.32        & 0.80/ 5.07       & 1.00/ 5.80       & 0.77/ 3.76        & 0.78/ 3.97       & 0.94/ 4.87       & 0.87/ 5.76        \\ \cline{2-14} 
                                     & \textbf{HWFET}                  & 0.88/ 3.01       & 1.39/ 5.50       & 1.20/ 5.41        & 0.55/ 3.40       & 1.10/ 5.27       & 0.64/ 2.89        & 1.04/ 7.10       & 1.45/ 7.64       & 0.96/ 6.55        & 0.9/ 4.57        & 1.30/ 5.43       & 1.50/ 5.88        \\ \hline
\end{tabular}
}
\end{center}
\end{table*}

As can be observed the dense first architecture with two convolutional layers performs reasonably better than the other architectures and gives a minimum MAE error of $0.58 \%$ and a MAX error of $2.78\%$ at 25 deg C. A possible explanation for this could be that appropriately stacked convolutional layers are able to extract complex underlying features relevant to SoC estimation while discarding unwanted information. In the remaining part of the paper we shall refer to this architecture as the $2C-DF$ architecture. We will also use only the $2C-DF$ architecture for further experimentation. Notably, the performance of all the other architectures are also appreciable and all of them provide MAE errors of less than $2 \%$.

\subsection{Model Evaluation}\label{subsecexp:mod eval}
\begin{figure}[ht]
    \centering
    \captionsetup{justification=centering}
    \includegraphics[keepaspectratio, width=\linewidth]{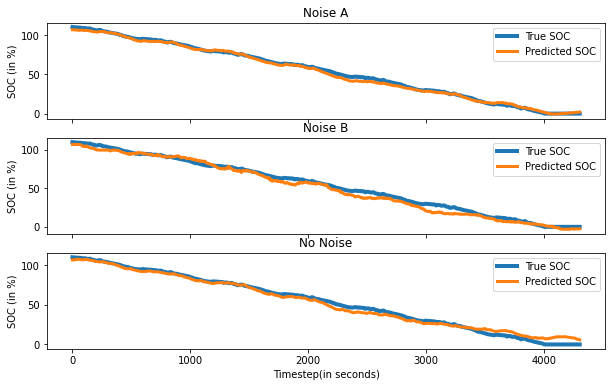}
    \caption{SoC estimation on US06 drive cycle at $25$\textdegree C from the Panasonic Dataset, sampled at 1 Hz, with different forms of noise. The CNN model consists of a 2C-DF architecture and $t_w$ was equal to 500.}
    \label{fig:Pan_SoC_final}
\end{figure}

\begin{figure}[!ht]
    \centering
    \captionsetup{justification=centering}
    \includegraphics[keepaspectratio, width=\linewidth]{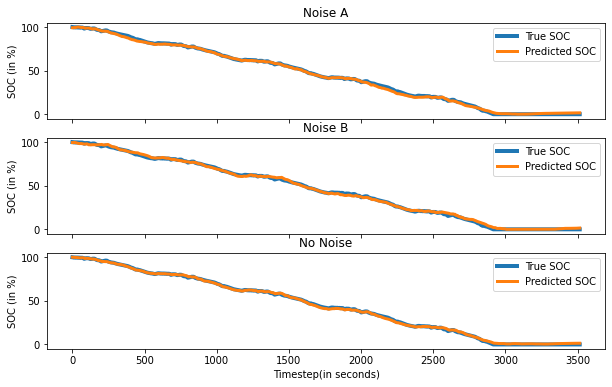}
    \caption{SoC estimation on US06 drive cycle at $25$\textdegree C from the LG Dataset, sampled at 1 Hz, with different forms of noise. The CNN model consists of a 2C-DF architecture and $t_w$ was equal to 500.}
    \label{fig:LG_SoC_final}
\end{figure}

The MAE and MAX estimation errors evaluated on the US06 and HWFET drive cycles of the Panasonic battery dataset, sampled at 10 Hz, 1 Hz and 0.1 Hz are plotted in figures \ref{fig:Pan_loss_var_original}, \ref{fig:Pan_loss_var_down_by_10} and \ref{fig:Pan_loss_var_down_by_100}\footnotemark respectively. Similarly, the MAE and MAX estimation errors evaluated on the US06 and HWFET drive cycles of the LG battery dataset, sampled at 10 Hz, 1 Hz and 0.1 Hz are plotted in figures \ref{fig:LG_loss_var_original}, \ref{fig:LG_loss_var_down_by_10} and \ref{fig:LG_loss_var_down_by_100}\footnotemark[\value{footnote}] respectively. In each of the subplots in these figures there are three plots corresponding to the estimation error of the model trained on data with Noise A, Noise B and with no noise. The errors are plotted as a function of $t_w$. The actual error values for $t_w$=\{100, 500, 1000\} can be found for the Panasonic dataset in table \ref{tab:tablePan}\footnotemark[\value{footnote}] and the LG dataset in table \ref{tab:tableLG}\footnotemark[\value{footnote}]. The predicted and true SoC values corresponding to the US06 drive cycle sampled at 1 Hz under an ambient temperature of 25 \textdegree C has been plotted in figure \ref{fig:Pan_SoC_final} for the Panasonic dataset and in figure \ref{fig:LG_SoC_final} for the LG dataset.\\
{}

{}

\begin{figure}[ht]
    \centering
    \captionsetup{justification=centering}
    \includegraphics[keepaspectratio, width=0.9\columnwidth]{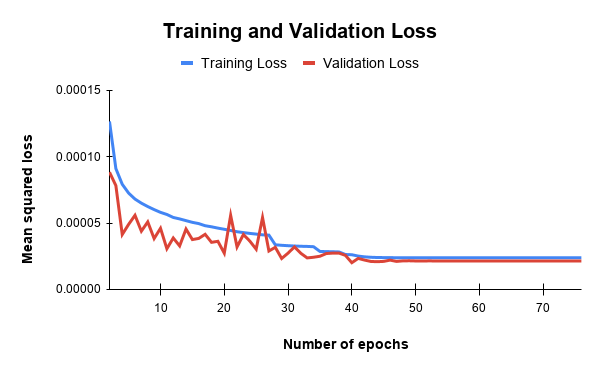}
    \caption{Training and Validation loss vs Epochs on Panasonic dataset, sampled at 10 Hz and with $t_w$ = 500 and no added noise.}
    \label{fig:train_vs_val_loss_Pan}
\end{figure}{}

\footnotetext{An unfortunate point to be noted here is that in the process of converting the original datasets, sampled at 10 Hz to datasets sampled at 0.1 Hz, a downsampling operation by a factor of 100 had to be performed on the original datasets, thus decreasing the number of sample data points by a factor of 100. As a result, the number of test data points available in US06 and HWFET drive cycles were lesser than and insufficient to carry out testing with $t_w$ =1000.  Hence, the results corresponding to data sampled at 0.1 Hz and $t_w$ =1000 have been omitted from the tables \ref{tab:tablePan} and \ref{tab:tableLG} and plots in figures \ref{fig:Pan_loss_var_down_by_100} and \ref{fig:LG_loss_var_down_by_100}. Also, with data sampled at 0.1 Hz and $t_w$ = 500, sufficient data was not available to carry out reliable testing for the individual test drive cycles and hence the combined test result of both the US06 and HWFET drive cycles have been presented in the aforementioned tables and plots.}

From the comparison plots and tables described above, the following inferences can be drawn:\\
\begin{enumerate}
    \item With an increase in $t_w$ both the MAX and MAE errors decrease drastically in the beginning and then after 500, the change is saturated. This happens because with small values of $t_w$, the model does not get to see much of the historical data at once to get enough information to accurately estimate the current SoC. In most cases, the results obtained with $t_w$ =1000 is marginally better than those obtained with $t_w$=500. However, to obtain this marginal increase in accuracy, a much heavier computational burden needs to be undertaken as we move from $t_w$ = 500 to 1000. Hence, in terms of computational expense and accuracy, the choice of $t_w$ = 500 has been found to be the most optimal one amongst the values tested.
    \item Another interesting phenomenon that can be pointed out from these results is that in most cases the estimation errors in case of data with added Noise A is lesser than data with no noise in it. The reason behind it is that the presence of white gaussian noise in training data fed to a deep learning model acts as an L2 regularizer for the weights of that model with a regularization coefficient proportional to the standard deviation of the noise \cite{10.5555/1162264}, thus further improving the generalizability of the model by reducing variance in its predictions and preventing it from overfitting.
    \item Compared to Noise A and the no noise scenario, the model finds it comparatively difficult to fit to the data with added Noise B, which is of a high value and non-gaussian nature. However, the estimation errors are still quite low in this case implying that the regularization techniques implemented during training makes the proposed SoC estimator robust to noises in data and prevents degradation in performance even under complex and high noise situations.
    \item Compared to the estimators proposed or mentioned in \cite{CHEMALI2018242}, \cite{8240689}, \cite{Li_2019} and \cite{9036949}, our proposed estimator gives lesser or comparable estimation errors measured on the same dataset. Moreover the number of epochs required for training the LSTM model in \cite{8240689} and the GRU model in \cite{Li_2019} is considerably higher and in the range of a few thousands.
\end{enumerate}

%An unfortunate point to be noted here is that in order to convert the original datasets which were sampled at 10 Hz to datasets sampled at 0.1 Hz, the data had to be downsampled by a factor 100. As a result, the number of sample data points decreased by a factor of 100. As a result, in the scenario of $t_w$ = 1000, the total number of test data points available in combined US06 and HWFET drive cycles were insufficient to carry out the aforementioned testing procedures. Hence, the portions with data sampled at 0.1 Hz and $t_w$ =1000 have been left unoccupied in the tables \ref{tab:tablePan} and \ref{tab:tableLG} and omitted from plots in figures \ref{fig:Pan_loss_var_down_by_100} and \ref{fig:LG_loss_var_down_by_100}. Also, with data sampled at 0.1 Hz and $t_w$ = 500, sufficient data was not available to carry out testing for the individual test drive cycles and hence the combined test result of both the US06 and HWFET drive cycles have been presented in the tables and plots.

\begin{figure}[!ht]
    \centering
    \captionsetup{justification=centering}
    \includegraphics[keepaspectratio, width=\linewidth]{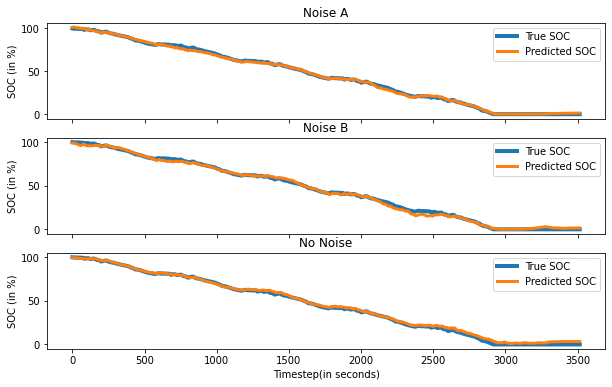}
    \caption{Transfer Learning Results: SoC estimation on US06 drive cycle at $25$\textdegree C from the LG Dataset, sampled at 1 Hz, with different forms of noise. The CNN model, trained using transfer learning with the entire LG training dataset, consists of a 2C-DF architecture and $t_w$ was equal to 500.}
    \label{fig:LG_TL_SoC_final}
\end{figure}{}

\begin{figure}[!ht]
    \centering
    \captionsetup{justification=centering}
    \includegraphics[keepaspectratio, width=\linewidth]{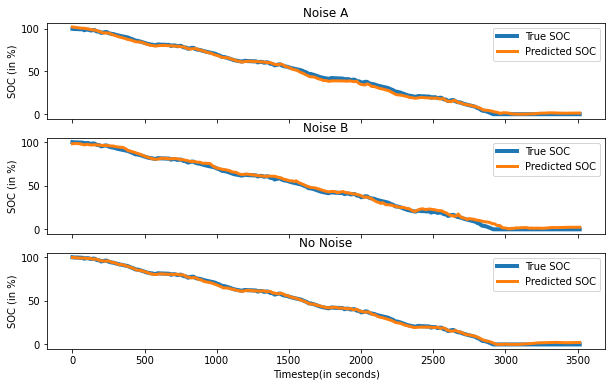}
    \caption{Transfer Learning Results: SoC estimation on US06 drive cycle at $25$\textdegree C from the LG Dataset, sampled at 1 Hz, with different forms of noise. The CNN model, trained using transfer learning with $40\%$ of LG training dataset, consists of a 2C-DF architecture and $t_w$ was equal to 500.}
    \label{fig:LG_TL_ig_60_SoC_final}
\end{figure}{}
\vspace{-5pt}

\begin{figure}[!ht]
    \centering
    \captionsetup{justification=centering}
    \includegraphics[keepaspectratio, width=\linewidth]{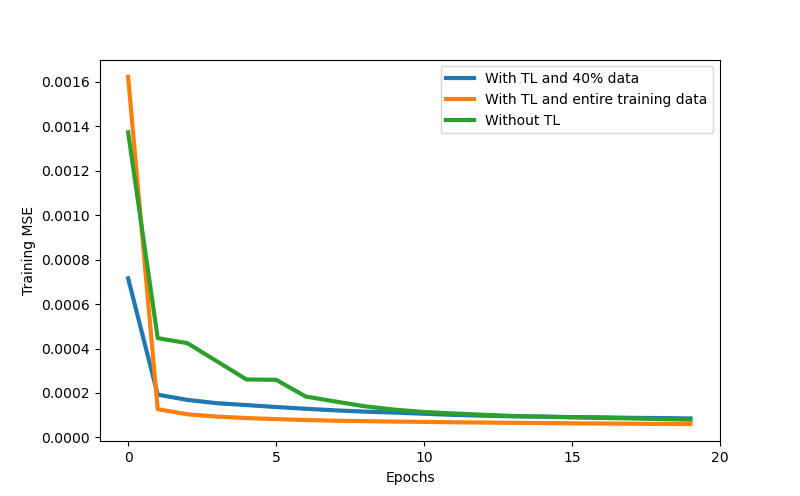}
    \caption{Training loss convergence characteristics for a noiseless LG dataset sampled at 10 Hz and with $t_w$ =500. Three plots correspond to cases when the CNN model is trained with and without transfer learning. The with transfer learning case is further of two types, one where the entire LG training data is used for training and the other where 40\% data is used for training.} 
    \label{fig:conv_char}
\end{figure}{}

\subsection{Transfer learning on different dataset}\label{subsecexp:TL}

The transfer learning strategy described in sections \ref{subsec:TL_intro} and \ref{subsec:TL_exp} and depicted in figure \ref{fig:CNN_training_TL} is followed to train the CNN model on the LG dataset. 
 MAE and MAX estimation of the transfer learning CNN model on the US06 and HWFET drive cycles of the LG dataset, sampled at 10 Hz and 1 Hz are plotted in the comparison plots in figures \ref{fig:LG_TL_loss_original} and \ref{fig:LG_TL_loss_down_by_10} respectively. The three plots in each subplot correspond to the three noise conditions namely, noise A, noise B and no noise, described in previous sections and the variations are measured across $t_w$ = \{100,500,1000\}. These plots correspond to the scenario where the transfer learning CNN model is trained on the entire LG training dataset. Similar plots corresponding to the scenario where the model is trained on stochastically chosen 40\% of the LG training drive cycles is given in figure \ref{fig:LG_loss_TL_ig_60}. In this scenario, the value of $t_w$ is fixed at 500. The actual estimation error values corresponding to each of these plots has been put down in table \ref{tab:tableLG_TL}. The predicted and true SoC values corresponding to the US06 drive cycle sampled at 1 Hz under an ambient temperature of 25\textdegree C has been plotted in figure \ref{fig:LG_TL_SoC_final} for the scenario where the entire LG training dataset was used for training and in figure \ref{fig:LG_TL_ig_60_SoC_final}  for the scenario where only 40\%  of training dataset was available.
The following are some of the key inferences that can be drawn from the aforementioned plots and tables:
\begin{enumerate}
\item The performance of a CNN model trained using transfer learning is at par and in some cases even better than an identical CNN model trained without transfer learning on similarly conditioned datasets.
\item From the comparison plots in figure \ref{fig:LG_loss_TL_ig_60} and the data in table \ref{tab:tableLG_TL} it can be seen that with transfer learning the model performs almost at par to other similar models even with a substantially lesser amount of data available for training. Due to the intelligent initialization provided by the transfer of weights from an already learned model on a different domain to an untrained one, it becomes possible for the untrained model to adapt to the current domain and reach the corresponding optima with substantially lesser amount of new information. As an analogy, “Mastering a country song would be comparatively easier for a pop singer than a complete rookie”. Thus transfer learning makes the model highly data efficient.
\item It can be seen from figure \ref{fig:LG_TL_loss_down_by_10}, that with transfer learning, the model performs worse in most cases with a $t_w$ of 1000 than with a $t_w$ of 500. A possible explanation for this behavior is that with transfer learning, the amount of information required by the model to predict accurately is substantially reduced. Hence even with smaller values of $t_w$, it can learn better and if a whole lot of extra information is passed onto it as is happening in the former case where $t_w$ =1000, the model has to put a lot of effort in filtering them to identify the relevant bits and hence starts performing slightly worse. It should however be noted that this happens when the data is sampled at 1 Hz and not when the data is sampled at 10 Hz. A possible explanation can be that in the former case, one timestep corresponds to a higher quantum of information than the latter as the changes in data dynamics are more smoothened out in the latter case.
\item  Also, since only a small segment of the weights, corresponding to the convolutional layers of the transfer learning CNN model is retrained following the transfer of weights during its initialization , the training time is also reduced and the model converges faster as can be seen from figure \ref{fig:conv_char}. The training loss converges faster when transfer learning is used.
\end{enumerate}

%\vspace{-5pt}
\section{Conclusion}\label{sec:conclusion}
A novel 1D CNN based state of charge estimator has been proposed in this paper. A number of experiments were conducted by varying the sampling frequency of the datasets, the history size parameter $t_w$, the noise content in the data and the type of initialization of the CNN SoC estimator and exclusive results and inferences have been put forth. The CNN based SoC estimator has fared well in all of them and has emerged as a robust, reliable and accurate SoC estimating algorithm. The estimation error values obtained in these experiments are lower than or at par with any state-of-the-art machine learning based SoC estimation algorithms tested on the same dataset and under similar training conditions and validation metrics. As a model free SoC estimation algorithm, the development and implementation of the proposed method required no special knowledge of battery chemistry and SoC estimation was carried out with a high accuracy using only observable historical data. Moreover a transfer learning framework was proposed which rendered the CNN model highly data efficient and generalizable across different battery chemistries.

%\section*{Acknowledgment}

%This research is sponsored by the Department of Science and Technology, India through project number DST/CERI/MI/SG/2017/080.

\printbibliography

\end{document}